\documentclass[a4paper]{jpconf}
\usepackage{color}
\usepackage{colortbl}
\usepackage[dvipsnames]{xcolor} 
\usepackage{graphicx}

\begin{document}
\title{Nuclear Spin Relaxation of Very Dilute $^3$He impurities in Solid $^4$He}
\author{S. S. Kim, C. Huan, L. Yin, J. S. Xia and  N. S. Sullivan}
\address{Department of Physics, University of Florida, Gainesville, Fl
32611,USA}
\author{ D. Candela}
\address{ Department of Physics, University of Massachusetts, Amherst, MA 01003,
USA}
\ead{\textcolor{blue}{sungkim@phys.ufl.edu}}

\begin{abstract}
We report measurements of the nuclear spin-lattice and spin-spin
relaxation times of very dilute $^3$He in solid $^4$He in the temperature range
$0.01 \leq T \leq 0.5$ K for densities where anomalies have been observed
in torsional oscillator and shear modulus measurements. We compare the results
with the values of the relaxation times reported by other observers for higher
concentrations and the theory of Landesman that takes into account the elastic
properties of the $^4$He lattice. A sharp increase in the magnitude
of the nuclear spin-lattice relaxation times compared to the the classical
Landesman theory is observed close to the temperatures where the torsional and
shear modulus anomalies are observed. The NMR results suggest that the tunneling
of $^3$He impurities in the atomic-scale elastic distortion is affected by the same
processes that give rise to the macroscopic elastic dissipation anomalies.
\end{abstract}
\section{Introduction}
Great interest has been generated by the discovery of non-classical rotational
inertia fractions (NCRIFs) in solid $^4$He by Kim and
Chan \cite{KimChanNature,KimChanScience} as this could be evidence of a
transition to a state supporting superflow in the solid at low
temperatures \cite{Leggett}. In addition measurements of the elastic properties
of the crystal have demonstrated the existence of a frequency dependent change
in the shear modulus \cite{Day_Beamish} and an enhanced dissipation peak \cite{
PhysRevLett.104.195301} with a temperature dependence that appears to mimic that
of the NCRIF anomaly. An anomalous peak is also seen in the sound
attenuation \cite{springerlink:10.1007/BF02396904}.
These results suggest that the dynamical properties of the
$^4$He lattice play an important role in these two anomalous responses and we
therefore carried out NMR experiments \cite{SSK-PRL2011,Chao} to study the
microscopic dynamics for samples at the very low $^3$He
concentrations ($x_3\leq 25$ ppm) for which the NCRIFs are still clearly
visible. 

Measurements of the nuclear spin-lattice relaxation times ($T_1$)
(Fig.\,\ref{fig:
T1peak}) revealed a well-defined peak in the relaxation time in a narrow
temperature
range comparable to that reported for the other two anomalies
\cite{SSK-PRL2011}.
A similar but less well defined anomaly is seen in measurements of the nuclear
spin-spin relaxation times ($T_2$). These deviations from the expected
temperature independent relaxation induced by the tunneling of the $^3$He atoms
through the $^4$He lattice are distinct from the 
phase separation into
pure $^3$He droplets that occurs at lower temperatures.
In order to better understand the tunneling in the $^4$He lattice, the results at high temperatures ($T\geq0.2$~K) are 
compared with the values reported elsewhere for higher concentrations ($100\leq x_3\leq10,000$~ppm) for molar volumes
 $28.9 \leq V_m \leq 29.2$~cm$^3$ in the same temperature range $0.2\leq T\leq0.5$~K.
\begin{figure}[ht]
\hspace{1pc}
\begin{minipage}{16pc}
\includegraphics[%
width=0.98\linewidth,keepaspectratio]{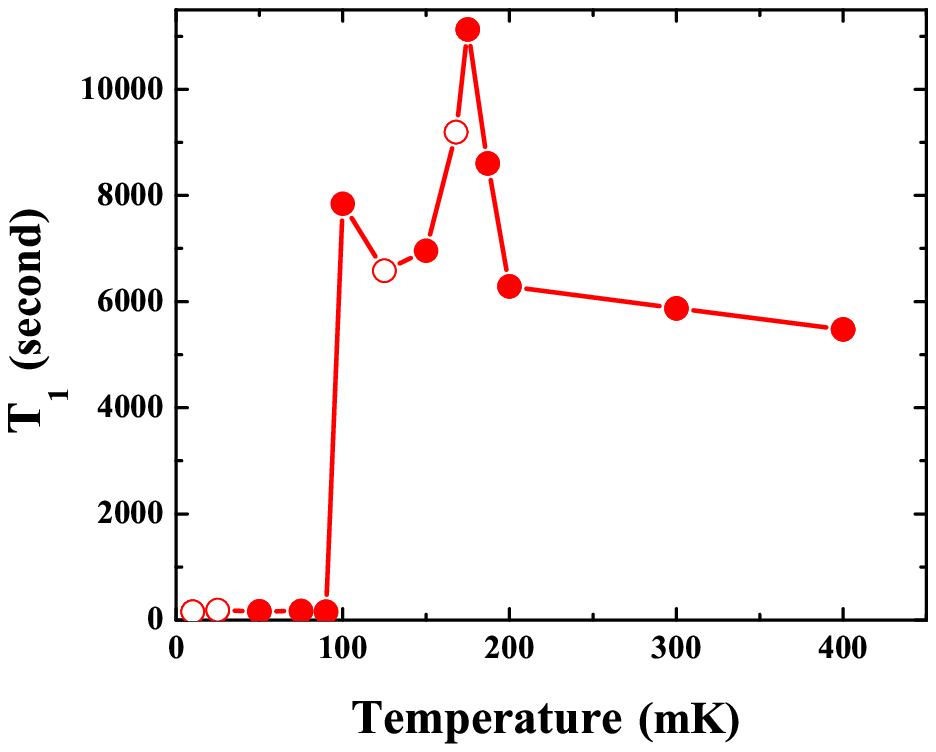}
\caption{\label{fig: T1peak} Temperature dependence of the nuclear spin-lattice
relaxation for a $^3$He concentration of 16 ppm (data from \cite{SSK-PRL2011}). The sharp drop
at 85 mK marks the $^3$He-$^4$He phase separation, while the narrow peak at 175 mK is
the anomalous temperature dependence discussed in Sec. \ref {sec: The Anomalous Nuclear Spin Relaxation}.}
\end{minipage}\hspace{3pc}
\begin{minipage}{16pc}
\includegraphics[width=0.98\linewidth,keepaspectratio]{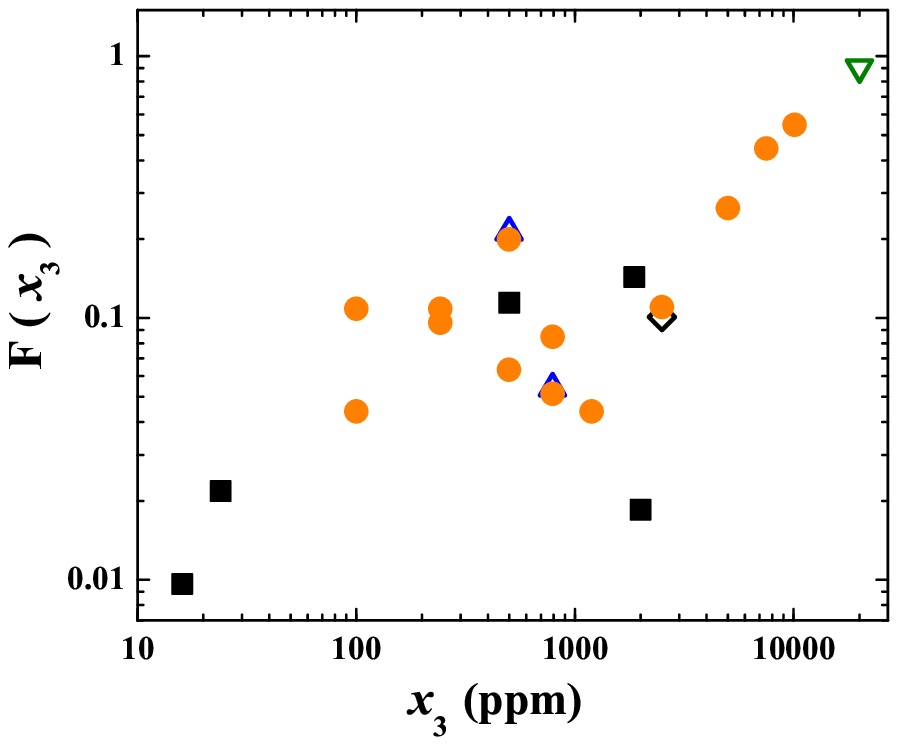}
\caption{\label{fig: T1T2}Variation of the function
$F(x_3)=({4\over 3})T_1T_2x_3^2M_2^2/\omega_L^2$ as a function of $^3$He
concentration. For a unique correlation time
$F(x_3)=1.0$. Experimental data: \fullsquare \,
Kim {\it et al.} \cite{SSK-PRL2011}, {\textcolor{blue} \opentriangle \, Allen {\it
et al.} \cite{Allen82}}, \textcolor{orange} \fullcircle \, 
Schratter {\it et al.} \cite{Schratter84},\, \opendiamond \, Schuster {\it et al.}
\cite{Schuster}, \textcolor{green} \opentriangledown \, Hirayoshi {\it et al.}
\cite{Hirayoshi}.}
\end{minipage} 
\end{figure} 
\section{Comparison of Results with Relaxation Models}
In order to understand the origin of the prominent peak in the relaxation time
at $T=175$ mK  we will first review the standard models of nuclear spin
relaxation expected for $^3$He impurities in otherwise pure solid
$^4$He and compare the results with published results for $16\leq x_3 \leq
10^4$ ppm \cite{SSK-PRL2011,Allen82, Schratter84, Schuster, Hirayoshi,
1972JLTP...8..3}.
The most remarkable result is that the data for $T_1$ and $T_2$
cannot be described by a common correlation time. In Fig.\,\ref {fig: T1T2} we
plot the product
$F(x_3)=({4\over 3})T_1T_2x^2_3M_2^2/\omega_L^2$ where $M_2$ is the NMR second
moment for pure $^3$He and $\omega_L$ is the Larmor frequency. For a simple
Lorentzian spectral density
with a unique correlation time, $F(x_3) =1$, independent of $x_3$. This is
clearly not satisfied. Attempts to use exponential spectral densities such as
that proposed for $(^3$He$)_2$ molecular motion \cite{Mullin75,Sacco76} also
fail. 
The latter leads to ratios $(T_1/T_2)\sim \exp(\omega_L/2J_{34})\sim3$ while the
experimental values
are $\sim 10^3$.
The results lead one to conclude that there are at least two components to the
spectral densities, one for the high frequency response and $T_1$, and a second
for the low frequency response and $T_2$.

Obvious candidates for the distinct components to the spectral densities are
(i) the group velocity of the tunneling $^3$He impurity, and (ii) the scattering
of that motion by the elastic fields of the lattice. Because of the difference
in 
zero-point motion of a $^3$He atom compared to a
$^4$He atom there is an appreciable lattice distortion surrounding the $^3$He
atoms described by a long range interaction $K(r_{ij})=
K_0/r_{ij}^3$ for $^3$He atoms at sites $i$ and $j$ \cite{Landesman, HuangPRB}. The
$^3$He motion is therefore determined by both the tunneling rate $J_{34}$ and
the elastic field constant 
$K_0$. The relaxation rates are determined by evaluating the expectation values
of the correlation functions $\Gamma_{ijkl}(t) = \langle
C_{ij}(t)C_{kl}(0)\rangle$ where
$C_{ij}(t)= \exp(-iK_it)[D_{ij},H_{34}]\exp(iK_jt)$
with $D_{ij}$ representing the tensorial components of the dipolar interactions.
$H_{34}$ is the $^3$He-$^4$He tunneling Hamiltonian. Two different approaches
have been made to calculate these expectation values: (1) the statistical method
of Landesman \cite{Landesman} valid for the high frequency limit of the spectrum
and estimations of $T_1$, and (2) the method of Huang {\it et al.}
\cite{HuangPRB} valid for $T_2$ calculations.
\begin{figure}[ht]
\hspace{1pc}
\begin{minipage}{16pc}
\includegraphics[width=18pc]{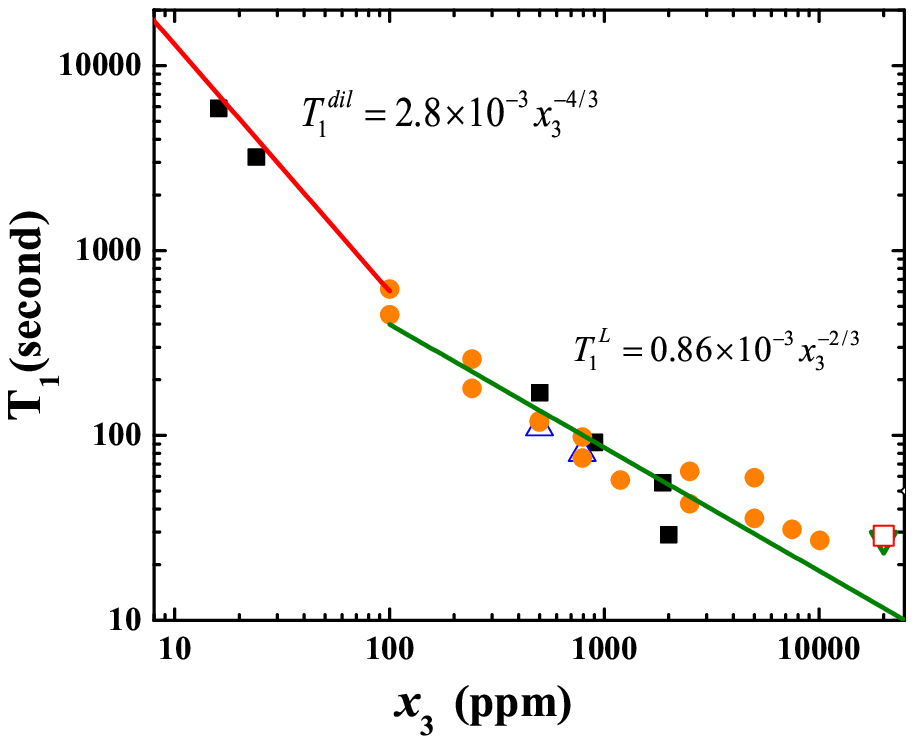}
\caption{\label{fig: CT1}Observed concentration dependence of the nuclear
spin-lattice relaxation times 
for dilute $^3$He in
solid $^4$He. \textcolor{red}\opensquare \, Greenberg {\it et al.}
\cite{1972JLTP...8..3} 
and for the rest of symbols see the caption for Fig. \ref {fig: T1T2}.}
\end{minipage}\hspace{3pc}
\begin{minipage}{16pc}
\includegraphics[width=18pc]{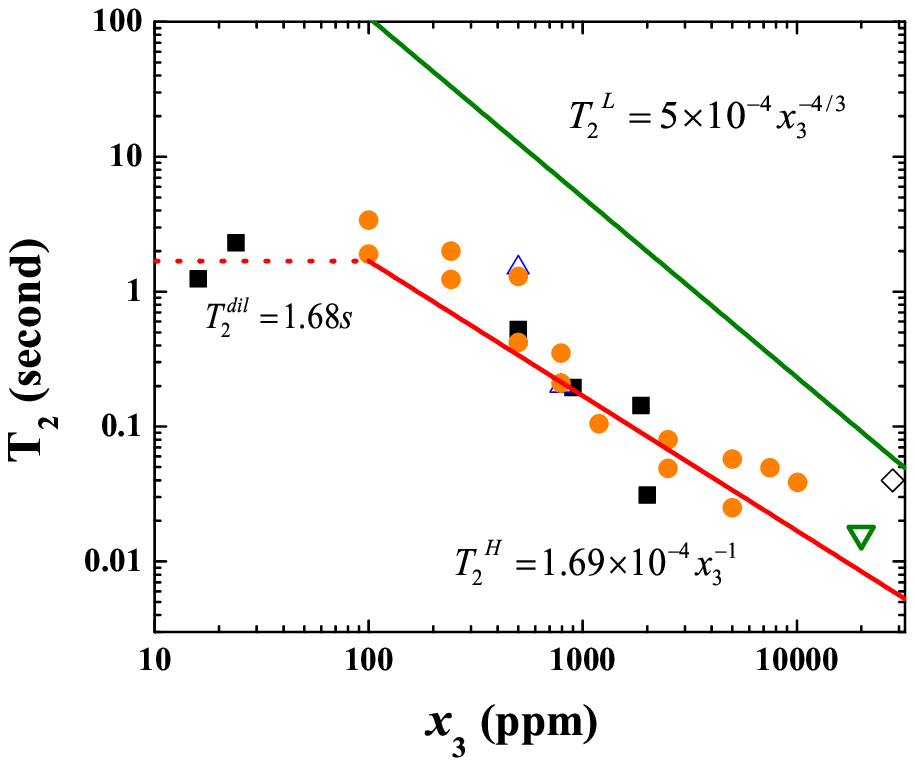}
\caption{\label{fig: CT2}Comparison of the concentration dependence of the
nuclear spin-spin relaxation times reported in the literature for dilute $^3$He in solid
$^4$He.   For symbols see the caption for Fig. \ref {fig: T1T2}.}
\end{minipage} 
\end{figure}
Landesman \cite{Landesman} evaluates the correlation function 
\begin{equation}\label{eqn:ALcorrel}
\Gamma(t) = 48J_{34}^2\sum_{i,j} (a_0/r_{ij})^6\prod_{p\neq(i,j)}
\exp[it(K_{ip}-K_{pj})]
\end{equation}
and, restricting the terms $K_{ip}-K_{pj}$ to values close to zero,
finds a correlation time $\tau_c^{-1}=23(J_{34}^2/K_0x_3^{1/3})$. For $\omega_L
> \tau_c^{-1}$, the calculated relaxation time is
\begin{equation}
T_1^{L}={{\omega_L^2\over M_2}{ K_0\over 46J_{34}^2}} = {0.86\times
10^{-3}x_3^{-2/3}}
\end{equation}
for $J_{34}/(2\pi)=1.8$ MHz and $J_{34}/K_0= 1.0 \times 10^{-3}$ chosen to
provide the best fit as shown by the solid green line in Fig. \ref {fig: CT1}. 
(The values of $T_1$ have been normalized to a Larmor frequency of
2 MHz, assuming a Lorentzian spectral density.)
For very dilute $^3$He concentrations, $x_3 \leq 200$ ppm, the $^3$He-$^3$He
collisions are too rare to be effective and the characteristic frequency that
determines the
relaxation is given by the time, $\tau_{ch}$, for the impurity to
travel the mean distance between sites. $\tau_{ch}^{-1} = x^{1/3}zJ_{34}$ where
$z$ is the lattice coordination number. This yields a relaxation time $T_1 =
2.8 \times 10^3x_3^{-4/3}$~s, shown by the solid red  line in Fig. 3.

If one uses the calculation of Landesman to estimate the spin-spin relaxation,
the result is off by an order of magnitude and predicts the incorrect
concentration dependence as shown by the  green line in Fig. 4. Huang {\it
et al.} \cite{HuangPRB} addressed this discrepancy by noting that
because of the $r^{-6}$ dependence in the correlation functions, only atoms very
close together can contribute  and that for an atom at site $r_i$ tunneling to
$r_j$ with a  spectator at $r_k$, only pairs for which $K(r_i)-K(r_j)= 2 J_{34}$
are to be considered. This energy conservation occurs for separations
$r\sim a_0(K_0/2J_{34})^{1/3}$ which becomes greater than the mean $^3$He
separation
for $x_3\leq 500$ ppm. Weighting the $r^{-6}$ sum in Eq. 1 by the number of
sites satisfying the conservation requirement, one estimates an effective
tunneling frequency $\tau^{-1}_H = 8.8 J_{34}^2/K_0 = 9.6 \times 10^4$ rad/s,
independent of $x_3$, using the same values of $J_{34}$ and $K_0$ as in
Landesman's calculation. The fit to the data  shown by the solid red line of
Fig. 4 is remarkably good given the approximations that have been made.
Huang {\it et al.} \cite{HuangPRB} carried out detailed Monte Carlo
calculations and showed that for $x_3\leq 400$ ppm
$\tau^{-1}_H$ develops a linear dependence on $x_3$, which leads to a
concentration and temperature independent $T_2$ as shown
by the dotted red line of Fig. \ref {fig: CT2}.
\section{The Anomalous Nuclear Spin Relaxation}\label{sec: The Anomalous Nuclear Spin Relaxation}
The approaches to the calculations of the nuclear spin 
relaxation
times of $^3$He impurities in solid $^4$He treat the distortion field around the
impurity as temperature independent. The experiments of Beamish {\it et al.}
\cite{Day_Beamish, PhysRevLett.104.195301} indicate that
there is \textquotedblleft in addition\textquotedblright a dynamical component that has a prominent temperature
dependence. Including a simple static thermal excitation
does not fit the data as it leads only to a
smooth step in the temperature dependence of $T_1$. Beamish {\it et al.}
\cite{Day_Beamish, PhysRevLett.104.195301}
introduced a dynamical relaxation $\tau(T)=\tau_0\exp(E/T)$ to describe the
frequency dependent shear modulus.  If we assume that the nuclear spin
relaxation occurs via this dynamical relaxation of the lattice in series with
the process described in section 2, then because the effective frequency of the
rapidly oscillating term in the relaxation function $\Gamma(t)$ is
$\omega_e=x_3^{4/3}K_0\sim 100$ Hz, the additional relaxation would be
expected to be proportional to the Lorentzian form 
$\omega_e\tau(T)/[1+(\omega_e\tau(T))^2]$ analogous to the anomalous term
observed for the shear modulus. While this temperature dependence does provide a
qualitative fit to the data (with an appropriate scale factor), a detailed
microscopic theory of the lattice relaxation is needed to justify the
introduction of this additional process for the relaxation.
\section{Conclusion}
The detailed concentration dependence of the nuclear spin relaxation of $^3$He
impurities in solid $^4$He for $10\leq T\leq 400$ mK can be understood if  we
assume that
 the tunneling of the $^3$He atoms is modified by the same process that leads to
the macroscopic dynamical relaxation observed in studies of the shear modulus.
The temperature dependence of this additional term fits a model of
a thermally activated relaxation for the lattice in response to the effective
motion of the tunneling in the local crystal fields.

\ack
The research was carried out at the NHMFL High B/T Facility which is supported
by NSF Grant DMR 0654118 and by the State of Florida.

\section*{References}

\bibliography{3Hein4He-SSK}

\end{document}